# *The FLUXCOM ensemble of global land-atmosphere energy fluxes*

Martin Jung[1], Sujan Koirala[1], Ulrich Weber[1], Kazuhito Ichii[2,3], Fabian Gans[1], Gustau-Camps-Valls[4], Dario Papale[5], Christopher Schwalm[6], Gianluca Tramontana[5], Markus Reichstein[1]

1. Max Planck Institute for Biogeochemistry, Hans-Knöll-Str. 10, 07745 Jena, Germany
2. Center for Environmental Remote Sensing, Chiba University, 1-33, Yayoi-cho, Inage-ku, Chiba, 236-0001, Japan
3. Center for Global Environmental Research, National Institute for Environmental Studies, 16-2, Onogawa, Tsukuba, 305-0053, Japan
4. Image Processing Laboratory (IPL). C/ Catedrático José Beltrán, 2. 46980 Paterna, Universitat de València, Spain
5. University of Tuscia DIBAF, Via C. de Lellis snc, 01100 Viterbo, Italy
6. Woods Hole Research Center, 149 Woods Hole Road, Falmouth, MA, 02540-1644, USA

corresponding author(s): Martin Jung (mjung@bgc-jena.mpg.de)

## Abstract

Although a key driver of Earth's climate system, global land-atmosphere energy fluxes are poorly constrained. Here we use machine learning to merge energy flux measurements from FLUXNET eddy covariance towers with remote sensing and meteorological data to estimate net radiation, latent and sensible heat and their uncertainties. The resulting FLUXCOM database comprises 147 global gridded products in two setups: (1) 0.0833° resolution using MODIS remote sensing data (RS) and (2) 0.5° resolution using remote sensing and meteorological data (RS+METEO). Within each setup we use a full factorial design across machine learning methods, forcing datasets and energy balance closure corrections. For RS and RS+METEO setups respectively, we estimate 2001-2013 global (± 1 standard deviation) net radiation as 75.8±1.4 W $m^{-2}$ and 77.6±2 W $m^{-2}$, sensible heat as 33±4 W $m^{-2}$ and 36±5 W $m^{-2}$, and evapotranspiration as 75.6±10 ×$10^3$ $km^3$ $yr^{-1}$ and 76±6 ×$10^3$ $km^3$ $yr^{-1}$. FLUXCOM products are suitable to quantify global land-atmosphere interactions and benchmark land surface model simulations.

## Background & Summary

Intercomparisons of global land surface models (LSMs) suggest large uncertainties regarding magnitude and pattern of land-atmosphere energy fluxes[1,2], making it difficult to assess and close energy and water budgets [3-5]. Existing regional networks of in-situ measurements from FLUXNET eddy covariance towers[6] provide only unevenly spaced point information impairing direct comparisons with LSMs. In addition, the lack of energy balance closure of circa 20% across sites suggests systematic biases of measured turbulent latent and sensible heat fluxes[7]. The reasons for the energy balance closure gap are unclear and a community-accepted correction is unavailable.

Previous efforts to integrate FLUXNET measurements, satellite remote sensing and climate data with machine learning[8-10] have yielded global products of land-atmosphere fluxes that have been used frequently to evaluate LSM simulations[11-15], for water budgets[16,17], and land-atmosphere interactions[18-21]. However, data-driven global flux estimates are subject to uncertainty from, for example, choice in machine learning algorithm and predictor variables, global climate forcing data, and the lack of energy balance closure. A better characterisation of these uncertainties is needed for energy and water budget studies and to interpret apparent mismatches with LSM simulations. This, in turn, will lead to improvements of global estimation of land-atmosphere energy fluxes by data-driven and process models.

The FLUXCOM initiative (www.fluxcom.org) aims to improve our understanding of the multiple sources and facets of uncertainties in empirical upscaling and, ultimately, to provide an ensemble of machine learning-based global flux products to the scientific community. We use two complementary experimental setups of input drivers (covariates) and resulting global gridded products. In the remote sensing ("RS") setup, fluxes are estimated exclusively from Moderate Resolution Imaging Spectroradiometer (MODIS) satellite data. The second approach additionally includes meteorological information. In this "RS+METEO" setup, fluxes are estimated from meteorological data and mean seasonal cycles of satellite data. Global products of the RS setup have the advantage that they do not require global climate forcing datasets as inputs. Such datasets are themselves subject to uncertainty and are limited in spatial resolution. Not using climate data however excludes potentially important information on meteorological conditions for biosphere-atmosphere fluxes and limits temporal coverage to the MODIS era (i.e. 2001 onwards). In contrast, the RS+METEO setup makes use of daily meteorological conditions and through the use of mean seasonal cycles of satellite derived input drivers allows for estimating fluxes beyond the satellite era.

The skill of machine learning-based estimation for both setups at flux tower sites was analysed in detail via cross-validation[22]. The analysis revealed good performance for latent and sensible heat, and in particular for net radiation. Both seasonality and between-site mean fluxes were well predicted--showing more skill than carbon fluxes (gross primary productivity, terrestrial ecosystem respiration, net ecosystem exchange). Furthermore, only negligible differences were found between different machine learning techniques, and between the RS and RS+METEO setups, which suggests an overall robust extraction of the main patterns of flux variation across methods.

In this study, we used the validated and trained machine learning techniques for the FLUXCOM energy fluxes of Tramontana et al.[22] and generated a large ensemble of gridded flux products. For the RS setup, nine machine learning methods were used to generate gridded products at an 8-daily temporal and 0.0833° spatial resolution for the 2001-2015 period. For the RS+METEO setup three machine learning techniques with four global climate forcing data sets yielded products with daily temporal and 0.5° spatial resolution and time periods (from ~1980 to present) depending on the climate input data. For latent and sensible heat fluxes, we additionally considered uncertainty from a lack of tower-based energy balance closure by propagating three different correction variants. Within the RS and RS+METEO setups, we followed a full factorial design of machine learning methods (9 for RS, 3 for RS+METEO) times energy balance correction variants (3 for LE and H, 1 for Rn), and climate forcing input products (4, only for RS+METEO). To allow for a better reuse of the large archive, we generated ensemble products by pooling machine learning estimates and energy balance closure gap variants. For the RS+METEO setup this was also done separately for each climate forcing data to allow modellers to compare their simulations with the FLUXCOM ensemble product driven by the same forcing.

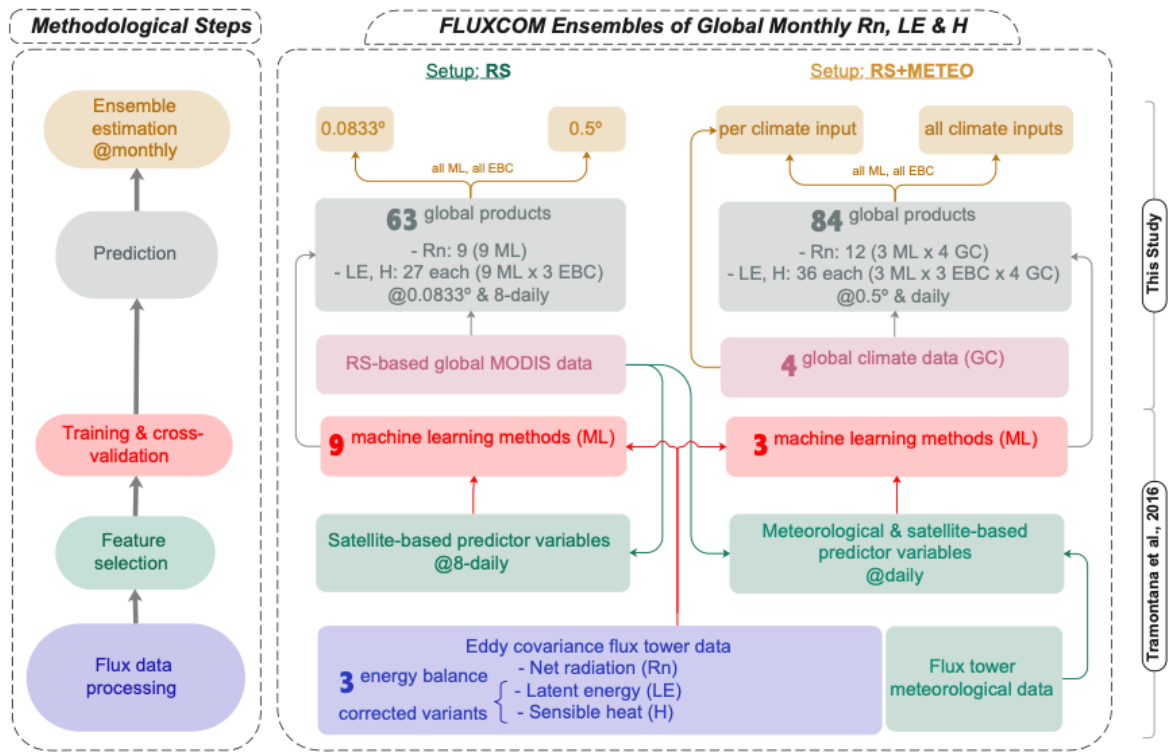

**Figure 1. Schematic overview of the methodology and data products from the FLUXCOM initiative.** The flow diagram shows the methodological steps for the remote sensing -based (RS, left) and the remote sensing and meteorological data -based (RS+METEO, right) FLUXCOM products. Final monthly ensemble products for Rn, LE, and H from RS are available at 0.0833° and at 0.5° spatial resolution. Ensemble products from RS+METEO are available per climate forcing data set as well as a pooled ensemble. All ensemble products encompass ensemble members of different machine learning methods (ML, 9 for RS, 3 for RS+METEO) and energy balance corrections (EBC, 3 for LE and H).

## Methods

**Training of machine learning algorithms**

Machine learning methods were trained using observations from 224 flux tower sites following the specifications in Table 1 and detailed previously[22]. The observed flux tower data had been screened for good quality by (1) allowing for no more than 20% of half-hourly data comprising a daily value to be gap-filled data, (2) checking for empirical consistency of energy fluxes, and (3) visual inspection. The same set of valid data points was used for net radiation, latent, and sensible heat flux (i.e. case-wise exclusion). Daily latent and sensible heat fluxes were then corrected for energy balance closure gap at FLUXNET sites using different approaches before training the machine learning algorithms (see below). The choice of satellite based and meteorological predictor variables followed a thorough feature selection analysis using a tailored genetic algorithm[23]. Some predictor variables vary only in space (e.g. plant functional type), some also seasonally (e.g. potential shortwave radiation), and some for each individual time step and location (e.g. short wave radiation, see Table 1). Each machine learning method used the same dataset for training within the RS or RS+METEO setup respectively and used all available data points in contrast to using only 90% of sites as in the cross-validation analysis[22].

|  | RS | RS+METEO |
|---|---|---|
| **Product specifications** | | |
| **Spatial resolution** | 0.0833° | 0.5° |
| **Temporal resolution** | 8 daily | daily |
| **Time period** | 2001-2015 | ~1980-present (Depending on climate forcing) |
| **Climate input** | n.a. | CRUNCEPv8, WFDEI, GSWP3, CERES-GPCP |
| **Tiling by PFT** | no | yes |
| **Spatial & Seasonal patterns** | f(RS) | f(RS,METEO) |
| **Interannual & trend patterns** | f(RS) | f(METEO) |
| **Training specifications** | | |
| **Machine learning methods** | 9: RF, ANN, MARS, MTE (3 variants), KRR, SVR | 3: RF, ANN, MARS |
| **Number of flux observations for training** | ~20,000 | ~200,000 |
| **Spatial features** | PFT, Max of MSC(fAPAR*Rg), Min of MSC(Rg) | PFT, Max of MSC($WAI_U$), Mean of MSC(BAND 6), Max of MSC(fPAR*Rg) |
| **Spatial, seasonal features** | Rpot, MSC(EVI*$LST_{Day}$) | Rpot, MSC(NDWI), MSC($LST_{Night}$), MSC(EVI*Rg) |
| **Spatial, seasonal, interannual features** | Rg, $LST_{Day}$, Anom of $LST_{Night}$, Anom of (EVI*$LST_{Day}$) | Rg, Rain, Rh, Rg*IWA*MSC(NDVI) |

Table 1: Specifications of the FLUXCOM RS and RS+METEO setups for energy fluxes. List of acronyms: Enhanced Vegetation Index (EVI), fraction of Absorbed Photosynthetically Active Radiation (fAPAR), Leaf Area Index (LAI), daytime Land Surface Temperature ($LST_{Day}$) and night time Land Surface Temperature ($LST_{Night}$), Middle Infrared Reflectance (band 7) ($MIR^{(1)}$), Normalized Difference Vegetation Index (NDVI), Normalized Difference Water Index (NDWI), Plant Functional Type (PFT), incoming global Radiation (Rg), top of atmosphere potential Radiation (Rpot), Index of Water Availability (IWA), Relative humidity (Rh), upper Water Availability Index WAI ($WAI_U$) (for details see Tramontana et al. (2016) supplementary material, Sect. S3), Mean Seasonal Cycle (MSC). Random forest (RF), Artificial Neural Network (ANN), Multivariate Adaptive Regression Splines (MARS), Model-Tree Ensemble (MTE), Kernel Ridge Regression (KRR), and Support Vector Regression (SVR).

**Correction for energy balance non-closure at FLUXNET sites prior to training**

We used three different approaches to address uncertainty due to the widely observed lack of energy balance closure at FLUXNET sites. The different correction approaches correspond to different hypothesis regarding the primary cause of the energy balance closure gap. The general form of the correction is $x_{LE}*LE+x_H*H=Rn-G$ where $x_{LE}$ and $x_H$ are the correction factors for latent and sensible heat, respectively. The perhaps most widely used approach is the Bowen ratio correction[24] which assumes that the ratio of sensible and latent heat flux is accurately measured and LE, and H are scaled with the same correction factor ($x_{LE\&H} := x_{LE} = x_H$) to force energy balance closure ($x_{LE\&H}=(Rn-G)/(LE+H)$). The "residual approach" allocates all missing energy to either LE ($LE_{RES}$ with = $x_{LE}=(Rn-G-H)/LE$, $x_H=1$) or H ($H_{RES}$ with $x_H=(Rn-G-LE)/H$, $x_{LE}=1$). The correction factors $x_{LE}$ and $x_H$ are estimated as the median of 30 daily values in a moving window. Median values within moving windows were chosen to minimize the impact of noise on x. Very small fluxes were not corrected (x=1) because x can take implausible values when the denominator approaches zero.

**Global products of predictor variables for RS products**

To produce spatio-temporal grids of energy fluxes, the trained machine learning algorithms require only spatio-temporal grids of input data. We used MODIS land products (collection 5; https://lpdaac.usgs.gov/) as input data for FLUXCOM. The MODIS products include daytime and nighttime land surface temperature (LST; MOD11A2[25]), land cover (MCD12Q1[26]), fraction of absorbed photosynthetically active radiation by a canopy (fPAR) (MOD15A2[27]), and bidirectional reflectance distribution function (BRDF)-corrected reflectances (MCD43B4[28]). Land cover data from 2001 to 2010 were processed to assign the majority land cover class in each 0.0833° grid for the whole period, i.e. land cover change was not considered. The LST, fPAR, and BRDF-corrected reflectances were provided with an 8-day temporal resolution. The BRDF-corrected reflectances were further converted to vegetation indices: the normalized difference vegetation index (NDVI), the enhanced vegetation index (EVI)[29], and the normalized difference water index (NDWI)[30].

The processing of the gridded remote sensing data followed the procedure done at flux site-level[22]. Poor quality data were filled to create continuous time-series data. For each time snapshot, bad quality data were identified for each 1km pixels by the MODIS quality assurance/quality criteria (QA/QC). If more than 25% of 1 km pixels within a 0.0833° grid cell had good quality, the mean of good quality pixels was taken. Otherwise, the value was estimated using the local mean seasonal cycle, i.e. the mean value of other years (2000-2014) with accepted quality for the same 8-day period was used.

**Solar Radiation**

For the RS product we used incoming surface shortwave radiation data of the Japan Aerospace eXploration Agency (JAXA) Satellite Monitoring for Environmental Studies (JASMES) product for 2001-2015 period (ftp://suzaku.eorc.jaxa.jp/pub/GLI/glical/Global_05km/repro_v6/). The products are derived from Terra MODIS data with a simple radiative transfer model[31]. The products were previously evaluated for three EC sites in Asia[32] and 20 EC sites in Alaska[33], and showed a good agreement with observations. Spatial and temporal averaging was conducted by converting the original 5 km grid to 0.0833° grids and daily to 8-day temporal resolution. Missing data in the original 5km data were replaced by mean daily values of available years.

**Global products of predictor variables for RS+METEO products**

Mean seasonal and mean annual characteristics of MODIS-based remotely sensed land surface variables (See Table 1 and Tramontana et al for details) were tiled by plant functional type, i.e. grids for each PFT containing the mean value per PFT and time step at 0.5° were created. Daily mean seasonal cycles of MODIS data for each grid cell used in the RS+METEO setup were computed by linearly interpolating a temporally smoothed 8-daily mean seasonal cycle. The land cover fractions are based on the same product and approach as in the RS product.

For daily meteorological variables four different commonly used global climate forcing data sets were chosen: WATCH Forcing Data ERA Interim (WFDEI[34], 1979-2013, ftp://rfdata:forceDATA@ftp.iiasa.ac.at), Global Soil Wetness Project 3 forcing (GSWP3, 1950-2014, Data Citation 1) http://search.diasjp.net/en/dataset/GSWP3_EXP1_Forcing), CRUNCEPv8[35] (1950-2016, https://vesg.ipsl.upmc.fr/thredds/catalog/work/p529viov/cruncep/V8_1901_2016/catalog.html), and a combination of radiation based on CERES[36] and precipitation from GPCP[37] (CERES-GPCP, 2001-2014, https://ceres.larc.nasa.gov/, https://precip.gsfc.nasa.gov/). The water availability index and the index of water availability, (WAI and IWA, see supplement 3 in Tramontana et al.[22]), were calculated for each forcing data set based on daily precipitation and potential evapotranspiration. The native spatial resolution of all four climate forcing datasets was 0.5° except for CERES-GPCP (1°). Here, CERES based radiation and GPCP based precipitation data were regridded to 0.5° by splitting up the original 1° grid cells into 0.5° grid cells.

**Generation of global products (Prediction)**

For the RS products, the trained machine learning models were applied to the gridded predictor variable fields for each 8-daily time step with a spatial resolution of 0.0833°. For the RS+METEO products, the trained machine learning models were run for each daily time step and for each plant functional type (PFT) at a 0.5° spatial resolution separately, and a weighted mean over the PFT fractions was obtained for each gridcell and time step. Note that the fraction of unvegetated (barren, permanent snow or ice, water) area was omitted in that calculation such that the definition of the calculated flux densities are per vegetated area (rather than grid cell or land area). The omission of deserts was necessary due to a lack of flux tower data. This complicates the assessment of globally integrated fluxes for sensible heat and net radiation where these fluxes typically show large positive and negative fluxes for hot and cold deserts, respectively. All computations were performed with MATLAB on a high-performance computing cluster at the Max Planck Institute for Biogeochemistry, Jena.

**Spatial and temporal aggregation of FLUXCOM-RS products**

To facilitate broader reuse of the FLUXCOM-RS products originally at 0.0833° and 8-daily time step we derived monthly products at 0.0833° and 0.5°. The monthly temporal aggregation is based on linearly interpolating the 8-daily data into daily data followed by calculating monthly averages. The spatial aggregation to 0.5° is based on taking the mean value of non-missing data points within each 0.5° cell. A 0.5° grid with the number of valid data points in the original RS product per 0.5° cell is provided in a separate data file.

**Ensemble estimates**

For the RS setup, we generate ensemble products at 0.0833° and 0.5° spatial resolution for each flux (LE, H, Rn) by pooling all the different runs per machine learning (9) method and energy balance correction variants (3, only for LE and H). This yields 27 ensemble members for LE and H, and 9 for Rn for the RS ensembles.

For the RS+METEO setup, we generated ensembles for each climate forcing data by pooling the runs for three machine learning methods and energy balance correction variants (3, only for LE and H). For each climate forcing specific ensemble, this yields 9 ensemble members for LE and H, and 3 ensemble members for Rn. We additionally generated an overall RS+METEO ensemble by pooling runs for different climate forcing data, machine learning methods, and energy balance correction variants. For the overall RS+METEO ensemble this yields 36 ensemble members for LE and H, and 12 for Rn.

The ensemble products were generated for mean monthly fluxes where the ensemble estimate is the median over ensemble members for each gridcell and month. In addition, we included the median absolute deviation as a robust estimate of ensemble spread, i.e., uncertainty. For the RS ensemble, the ensemble spread captures uncertainty related to the choice of machine learning method and the lack of energy balance closure seen in FLUXNET data. For the overall RS+METEO ensemble, the ensemble spread captures uncertainty related to the choice of machine learning method, the energy balance closure gap issue, and the choice of climate forcing data. Due to space restrictions and conciseness, results and technical validation (see below) focus on a parallel assessment of the RS and the RS+METEO ensemble. Occasionally we make use of all ensemble members where appropriate (see Figure captions).

**Cross-consistency checks with the state-of-the-art estimates**

We compare the spatial patterns of mean annual LE and Rn fluxes as well as monthly time series of their continental means from the FLUXCOM ensemble against previous estimates. For LE we compare FLUXCOM RS and RS+METEO ensembles against those from Model Tree Ensemble (MTE[10]), the Global Land Evaporation Amsterdam Model (GLEAM v3.1a[38]), and LandFlux-EVAL[39]. The MTE (https://www.bgc-jena.mpg.de/geodb/projects/Home.php) is based on only one machine learning method[8] trained on monthly flux data[9,10] and may be regarded as a precursor to FLUXCOM. GLEAM (https://www.gleam.eu/) models evapotranspiration based on a Priestley-Taylor formulation[40] with explicit soil moisture stress, and the interception by the Gash model[41], and was informed by various satellite forcing data. LandFlux-EVAL (http://www.iac.ethz.ch/group/land-climate-dynamics/research/landflux-eval.html) is the ensemble mean of 14 evapotranspiration products of different approaches. For the conversion between evapotranspiration and latent heat we assumed a constant latent heat of vaporization of 2.45 MJ mm$^{-1}$.

For Rn, we compare against two satellite based products from Clouds and the Earth's Radiant Energy System (CERES, https://ceres.larc.nasa.gov/) SYN1d Ed4A product[36]) and Surface Radiation Budget (SRB, https://eosweb.larc.nasa.gov/project/srb/srb_table) release 3.1. Both products combine diverse atmospheric satellite data extensively with data assimilation while SRB can be regarded as a precursor to CERES. For comparison with FLUXCOM, the original 3-hourly data were aggregated to monthly means and resampled from 1° to 0.5° using the nearest neighbour method.

The comparisons of all data products use the common 2001-2005 period and a common mask for vegetated land area. Differences in masking may lead to some differences in mean global numbers compared to those reported elsewhere. We compare mean annual fluxes across space (Figure 4 and 5) and provide Pearson's correlation coefficient, equations of a linear total least squares fit, and density scatter plots of mean annual fluxes using gridcells with a land fraction of at least 80% to minimize inconsistencies in cross-product comparisons. For continental mean monthly fluxes (Figure 6) we calculated the mean and median absolute deviation (MAD) across all ensemble members for each monthly time step. MAD was converted into a robust estimate of 1 standard deviation by multiplying it with 1.4826 (assuming a normal distribution). The calculation of global and continental mean annual energy fluxes and their uncertainty (Figure 7 and 8) also follows the procedure of first aggregating each ensemble member for the period 2001-2013 but the common mask of valid data from the intersection with independent products was not used.

For an unbiased comparison of FLUXCOM sensible heat and net radiation fluxes with global values from the literature, we scaled FLUXCOM products to incorporate fluxes from non-vegetated area of the world. The non-vegetated land area not covered by FLUXCOM products corresponds to cold (mainly Greenland and Antarctica) and hot (mainly Sahara) deserts. For hot deserts, we estimated a mean sensible heat flux based on CERES net radiation and GPCP precipitation assuming that all precipitation is converted to latent heat and subtracted from Rn. The average values were computed for grid cells where the fraction of hot desert exceeds 50% and resulted: 5.9356 MJ $m^{-2}$ $day^{-1}$ for Rn and 5.8264 MJ $m^{-2}$ $day^{-1}$ for H for the period 2001-2010. For cold deserts we obtained mean Rn as -0.1826 MJ $m^{-2}$ $day^{-1}$ from CERES, while H was derived by a previously calculated value[42] of -33.2 W $m^{-2}$ (-2.8685 MJ $m^{-2}$ $day^{-1}$) based on reanalysis. The global adjusted value of FLUXCOM for sensible heat or net radiation was then computed as a weighted average for the three area fractions: vegetated = 0.765, cold deserts = 0.108, hot deserts = 0.1265, where the vegetated value is directly from FLUXCOM.

**Code availability**

Python code to synthesise the results and to generate the figures 2 to 8 can be obtained through the public repository at https://git.bgc-jena.mpg.de/skoirala/fluxcom_ef_figures. MATLAB code for generating the flux products and ensemble estimates is available on request to Martin Jung (mjung@bgc-jena.mpg.de) for the sake of reproducibility. The collaborative nature of the FLUXCOM initiative and the demanding computing resulted in complex and large amounts of code that was customized to the HPC and file system of MPI-BGC and is therefore challenging to use. Code for processing MODIS satellite data is available on request to Kazuhito Ichii (ichii@chiba-u.jp).

## Data Records

The native FLUXCOM energy flux products amount to more than 4 TB of data. Products with daily or 8-daily temporal resolution or customized ensemble estimates are available on request to Martin Jung (mjung@bgc-jena.mpg.de). Monthly energy flux data of all ensemble members as well as the ensemble estimates from the FLUXCOM initiative (http://www.fluxcom.org) described here (DataCitation 2) are freely available (CC4.0 BY licence) from the data portal of Max Planck Institute for Biogeochemistry (https://www.bgc-jena.mpg.de/geodb/projects/Data.php) after registration. Choose 'FluxCom' in the dropdown menu of the database and select FileID 257. The users will be provided with an access to a ftp server. The ftp directory stores 214 GB of data and is structured in a consistent way with the file naming in Table 2. The folder structure was designed to facilitate easy download of relevant subsets of the archive and is as per the following convention:

<SETUP>/<TYPE>/<sReso> or <METEO>/<tRESO>

<SETUP> is either "RS" or "RS_METEO". <TYPE> is either "ensemble" or "member". At the third level, RS uses <sRESO> ("720_360" or "4320_2160") and RS+METEO uses <METEO> (see Table 2). <tRESO> is always "monthly" here.

The files are provided in network Common Data Form, version 4 (netCDF-4) data format (https://www.unidata.ucar.edu/software/netcdf/). The data files are named using the following convention:

<EF>.<SETUP>.<EBC>.<MLM>.<METEO>.<sRESO>.<tRESO>.<YYYY>.nc

The details of each of the <item> in the filenames are provided in Table 2.

| SN | item | information | prefix | values |
|---|---|---|---|---|
| 1 | <EF> | Energy flux | no prefix | - LE, H, Rn |
| 2 | <SETUP> | Upscaling set up | no prefix | - RS: for RS products<br>- RS_METEO: for RS+METEO products |
| 2 | <EBC> | Energy balance correction | EBC- | - ALL: for ensembles that include both energy balance corrected and uncorrected fluxes<br>- NONE: uncorrected energy fluxes<br>- BWR: energy fluxes corrected by Bowen's ratio method<br>- RES: energy fluxes corrected by residual method<br>Note that EBC is always NONE for Rn, because it was never corrected. |
| 3 | <MLM> | Machine learning method | MLM- | - ALL: for ensembles that include energy fluxes from all machine learning methods<br>- ANN, MARS, or RF for energy fluxes from RS+METEO<br>- Rfmiss, ANNnoPFT, MARSens, GMDH_CV, KRR, MTE, MTE_Viterbo, MTEM, or SVM for energy fluxes from RS |
| 4 | <METEO> | Meteorological data | METEO- | - ALL: for RS+METEO ensembles that include energy fluxes from all meteorological data<br>- CERES_GPCP, CRUNCEP_v8, GSWP3, or WFDEI: for RS+METEO products using different meteorological data<br>- NONE: for all energy fluxes from RS (because RS does not use meteorological forcing data) |
| 5 | <sRESO> | Spatial resolution | no prefix | - 720_360: for 0.5°x0.5° native RS+METEO and spatially aggregated RS data<br>- 4320_2160: for 0.0833°x0.0833° native RS data |
| 6 | <tRESO> | Temporal resolution | no prefix | - monthly: for all data files |
| 7 | <YYYY> | Year | no prefix | - the year for which the data is |

Table 2: Key to naming convention used for the folder structure and naming conventions.

For example, the file "LE.RS.EBC-ALL.MLM-ALL.METEO-NONE.720_360.monthly.2001.nc" is the RS-based ensemble latent heat energy of year 2001 that includes all energy balance corrected and uncorrected fluxes produced using all machine learning methods and no meteorological data aggregated to 0.5° spatial resolution (size 720 along longitude and 360 along latitude) and monthly temporal resolution.

For all types (ensemble or member) of data, the variable names for latent heat energy, sensible heat, and net radiation are LE, H and Rn, respectively. The data files for both RS and RS+METEO -based ensembles include additional variables with suffix '_MAD' (e.g., LE_MAD). This variable provides the data for uncertainty (median absolute deviation) among different ensemble members for each grid cell and month. The '_MAD' uncertainty variable is also in the same time, latitude, longitude coordinates. The ensembles from RS+METEO products, that include runs with four different forcing inputs, also have a variable with '_n' (e.g., LE_n in latitude, longitude coordinates). This variable stores the number of ensemble members included while calculating the median. The number of ensemble members varies in space in RS+METEO ensembles, because of the difference in land-sea mask of different meteorological data used. The data for each variable is defined by 3 dimensions: 'latitude' for latitudes and 'longitude' for longitudes in space, and 'time' for time. The data variables are defined by time, latitude, longitude coordinates. The header of the netCDF-4 data files includes the global attributes that lists the product (RS+METEO or RS), type of the data (either member or ensemble), the machine learning method(s), and meteorological data used (as per Table 2).

## Technical Validation

In this section, we present the main spatiotemporal features of the FLUXCOM energy flux products. We validate patterns and magnitudes of fluxes against previous state-of-the-art estimates, and expectations from theory and literature.

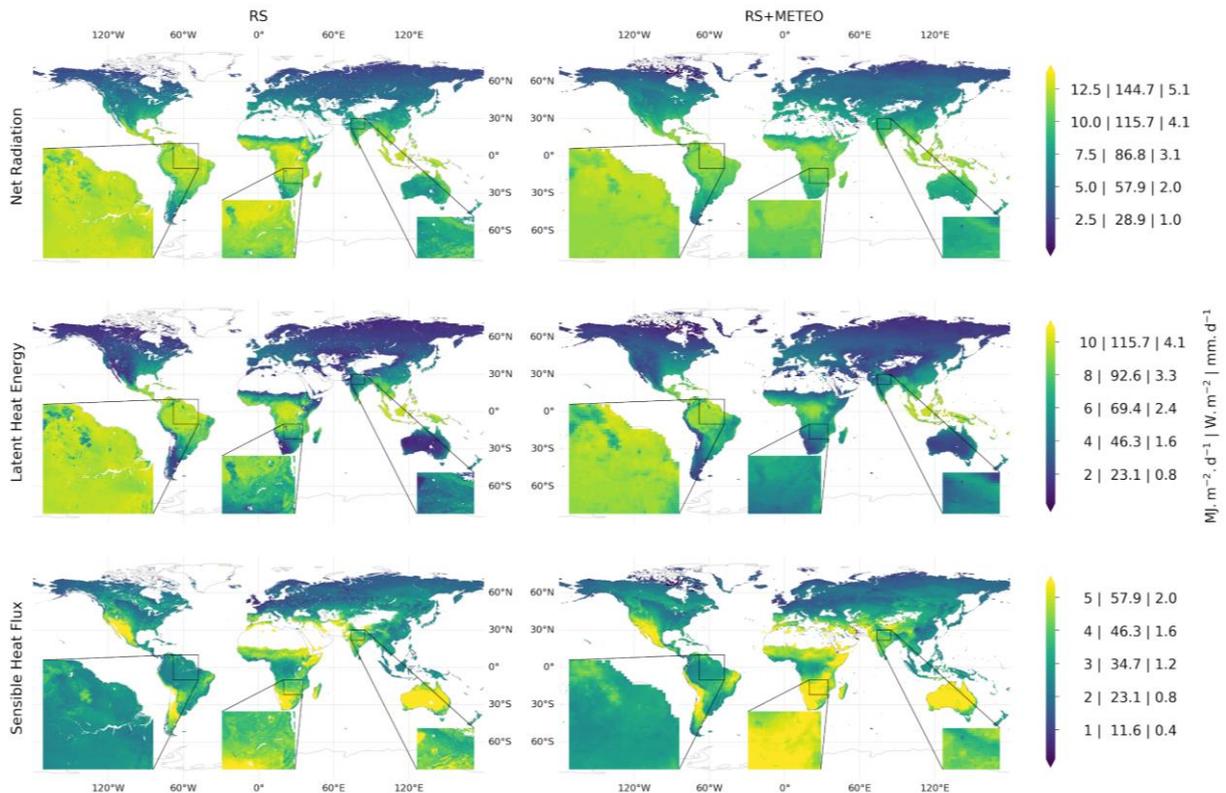

**Figure 2. Global distributions of mean annual (2001-2013) energy fluxes from the FLUXCOM RS and RS+METEO ensembles.** The first, second, and third row show net radiation, latent heat, and sensible heat fluxes, respectively (left: RS, right: RS+METEO). The fluxes are expressed in $MJ\ m^{-2}\ d^{-1}$, $W\ m^{-2}$, and $mm\ d^{-1}$, separated by '|', in the color bars. Inset figures show zooms for different regions.

Consistent with current understanding, mean annual latent heat and net radiation fluxes are the highest in tropical and the lowest in high latitude regions of the world (Figure 2). In contrast, mean annual sensible heat peaks in dry sub-tropical regions where latent heat fluxes are reduced due to expected water limitation on evapotranspiration which is known to increase the Bowen ratio (H/LE). These patterns are qualitatively consistent among the RS and RS+METEO ensemble products, while flux magnitude differences, e.g. larger net radiation of the RS product in the tropics, are also evident. In general, both RS and RS+METEO products show similar large-scale variations in energy fluxes but local-scale heterogeneities are better resolved in RS products (see the inset zoom-ins in Figure 2) due to a 6-fold increase in spatial resolution.

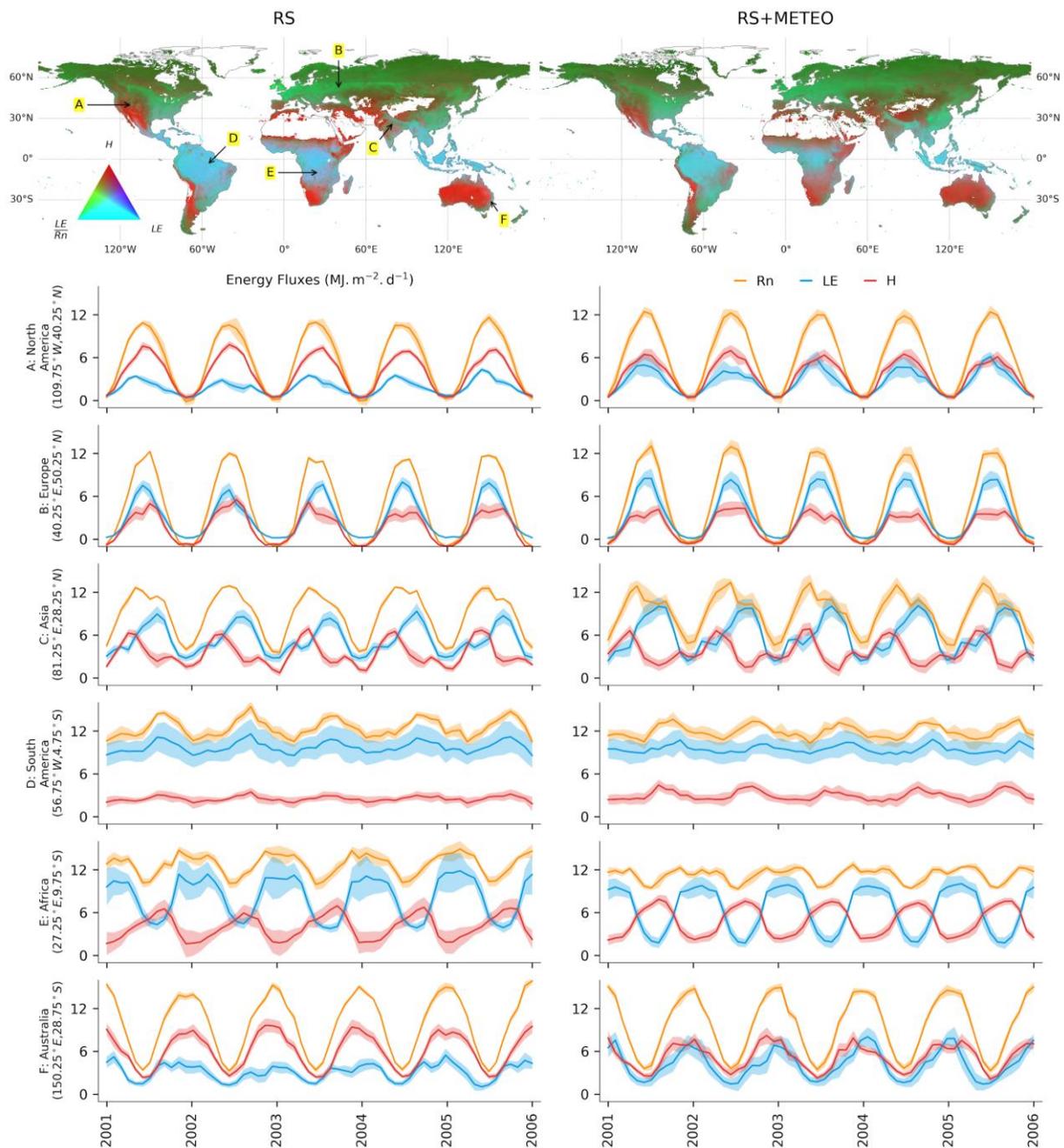

**Figure 3. Global covariations of land-atmosphere energy fluxes, and their temporal variations and uncertainties in selected locations.** RGB composite maps are with latent heat (*LE*) in the blue, sensible heat flux (*H*) in the red, and evaporative fraction (LE/Rn) in the green channel. Line plots show time series of *LE* (blue), *H* (red), and *Rn* (orange) for selected locations (0.5° grid cells, see map for RS) from 2001 to 2005 in *MJ $m^{-2}$ $d^{-1}$*. The shades around the lines indicate the uncertainty ranges (± 1 robust standard deviation) of ensemble members.

The top panels of Figure 3 provide a visual impression of the global spatial co-variation of energy fluxes. In these RGB maps, hot and dry regions appear as red where latent heat is low and net radiation is preferentially converted to sensible heat. Wet tropical regions with high net radiation and latent heat but low sensible heat appear as cyan. Regions where latent heat is energy limited but net radiation is intermediate or low appear in green. The partitioning of Rn into LE and H components is similar for both RS+METEO and RS products (Figure 3) with some regional differences visible. To illustrate local differences among the RS and RS+METEO ensemble products, as well as seasonal variations of energy fluxes and its uncertainties, we present time series of selected locations (0.5° grid cells) in subpanels of Figure 3. For example, in the selected location in North America (A), situated at the transition between water limited and energy limited regime of evapotranspiration, we see more H relative to LE in the RS ensemble compared to the RS+METEO ensemble. Similar patterns of slightly different net radiation partitioning in LE and H are also evident in other transitional locations in Africa (E) and Australia (F). Energy flux uncertainties (see shading in Figure 3) vary spatially, seasonally, and interannually as well as between the RS and the RS+METEO ensemble products. Where uncertainties of the RS+METEO ensemble are larger compared to the RS ensemble suggests larger contributions of meteorological forcing data uncertainty, while larger uncertainty of the RS ensemble may indicate larger contribution of machine learning method choice, perhaps due to poor constraints by flux tower stations. Overall, there is high level of consistency between the RS and RS+METEO ensemble products for seasonality and flux magnitudes as well as their uncertainties.

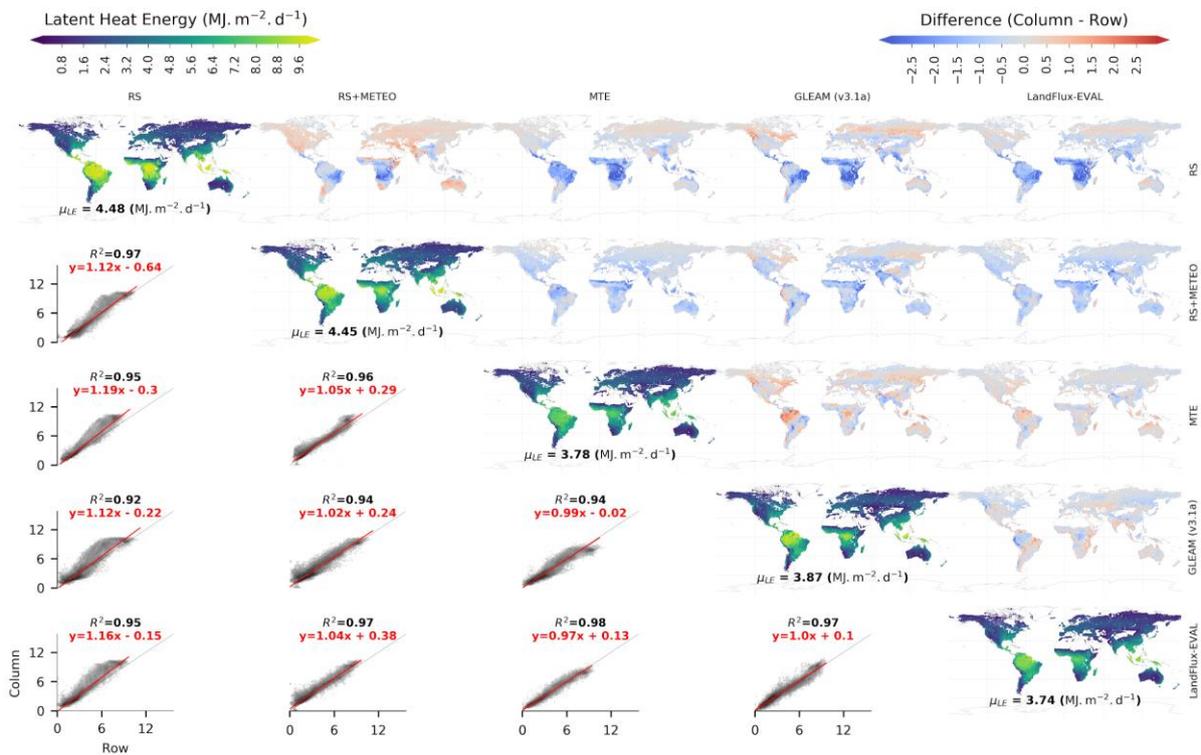

**Figure 4. Comparison of the global distributions of mean latent heat (LE) fluxes from FLUXCOM against previous global estimates.** Along the diagonal, maps of means for the period 2001-2005 from RS+METEO, RS, MTE, GLEAM v 3.1a, and LandFlux-EVAL are plotted along with the area weighted mean LE ($\mu_{LE}$) as text. Above the diagonal, the difference maps for each of these products (column - row) are plotted. Below the diagonal, density scatter plots between these products are provided with darker shade indicating larger density of points. Here, black lines show the 1:1 line, red lines show the total (orthogonal) least square regression fit with the equation given in red text and the squared Pearson's correlation coefficient ($R^2$).

We further evaluate the spatial patterns of mean annual energy fluxes of the FLUXCOM ensemble against previous estimates. First, we evaluate the long-term mean LE from RS and RS+METEO against those from MTE[10], GLEAM v 3.1a[38], and LandFlux-EVAL[39]. Generally, the spatial variation of ET is very consistent between FLUXCOM products and previous global estimates (Figure 4) with $R^2$ values close to 1. All show the dominant gradient between the highest LE in the humid tropics and the lowest LE in cold and dry places. There are however sizeable systematic differences between products, in particular within the tropics. Both RS and RS+METEO show larger LE in the tropics than MTE and LandFlux-EVAL while GLEAM shows regionally similar LE magnitudes in the wet tropics. The larger tropical LE in FLUXCOM propagates to 15-20% larger global means of LE compared to the other two estimates (see below for a broader comparison of global LE estimates). Due to the large LE flux in the wet tropics, even a comparatively small relative difference results in large absolute differences. In the tropical regions, the difference between RS and RS+METEO LE is also relatively large with larger LE in RS. But, globally, this difference is somewhat balanced by lower LE in RS in other regions. Semi-arid regions also tend to show comparatively large systematic differences across products.

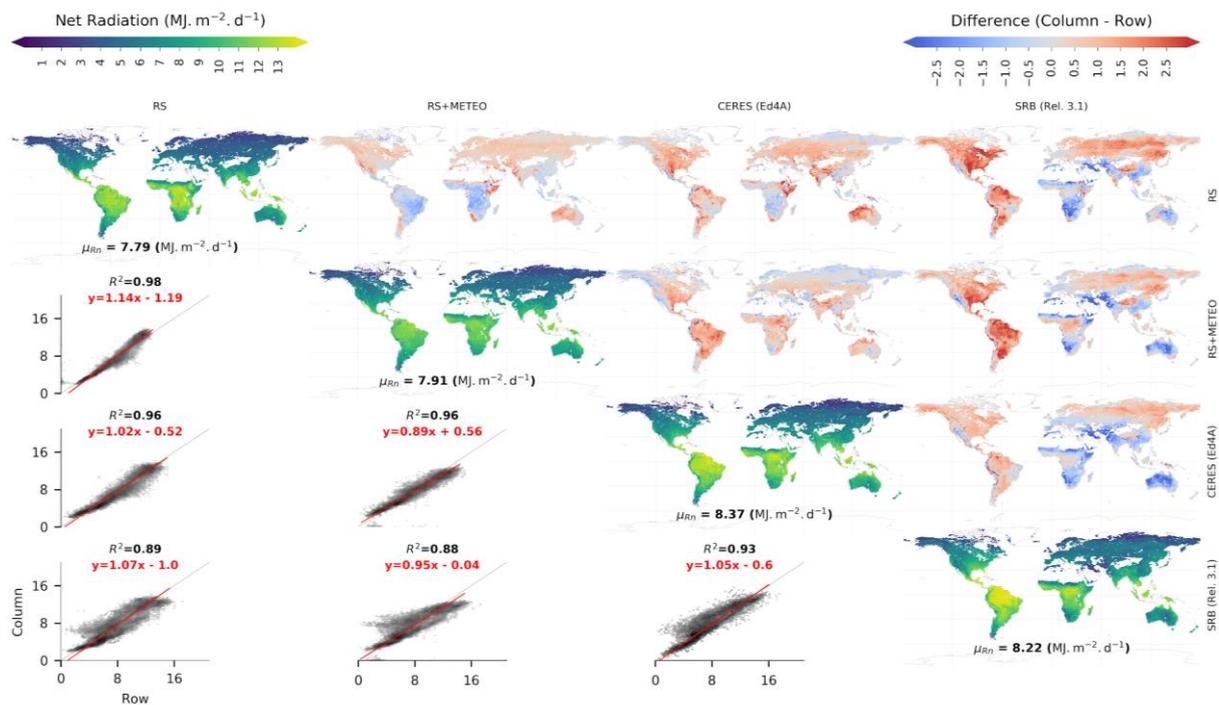

**Figure 5. Comparison of the global distributions of mean net radiation (Rn) from FLUXCOM against previous global estimates.** Along the diagonal, maps of the means for the period 2001-2005 from RS+METEO, RS, CERES, and SRB are plotted. See Figure 4 for further explanations. Note that the color scale of the difference maps is the same as for the LE (Figure 4) despite the larger flux magnitudes of Rn.

For Rn, we compare the spatial patterns of FLUXCOM products against two satellite based products from CERES[36] and SRB. The spatial patterns of mean annual Rn from RS and RS+METEO agree better with CERES ($R^2$=0.96) than the agreement between CERES and SRB ($R^2$=0.93) (Figure 5). The RS product and CERES show larger Rn in the tropics compared to RS+METEO. Large differences between SRB and all other products are evident both in tropical and extratropical regions, South America, large parts of North America and across Eurasia. CERES as well as SRB tend to show larger Rn in many extratropical regions compared to both FLUXCOM ensemble products. This contributes to a 4-7% larger global vegetated Rn of CERES and SRB compared to FLUXCOM products.

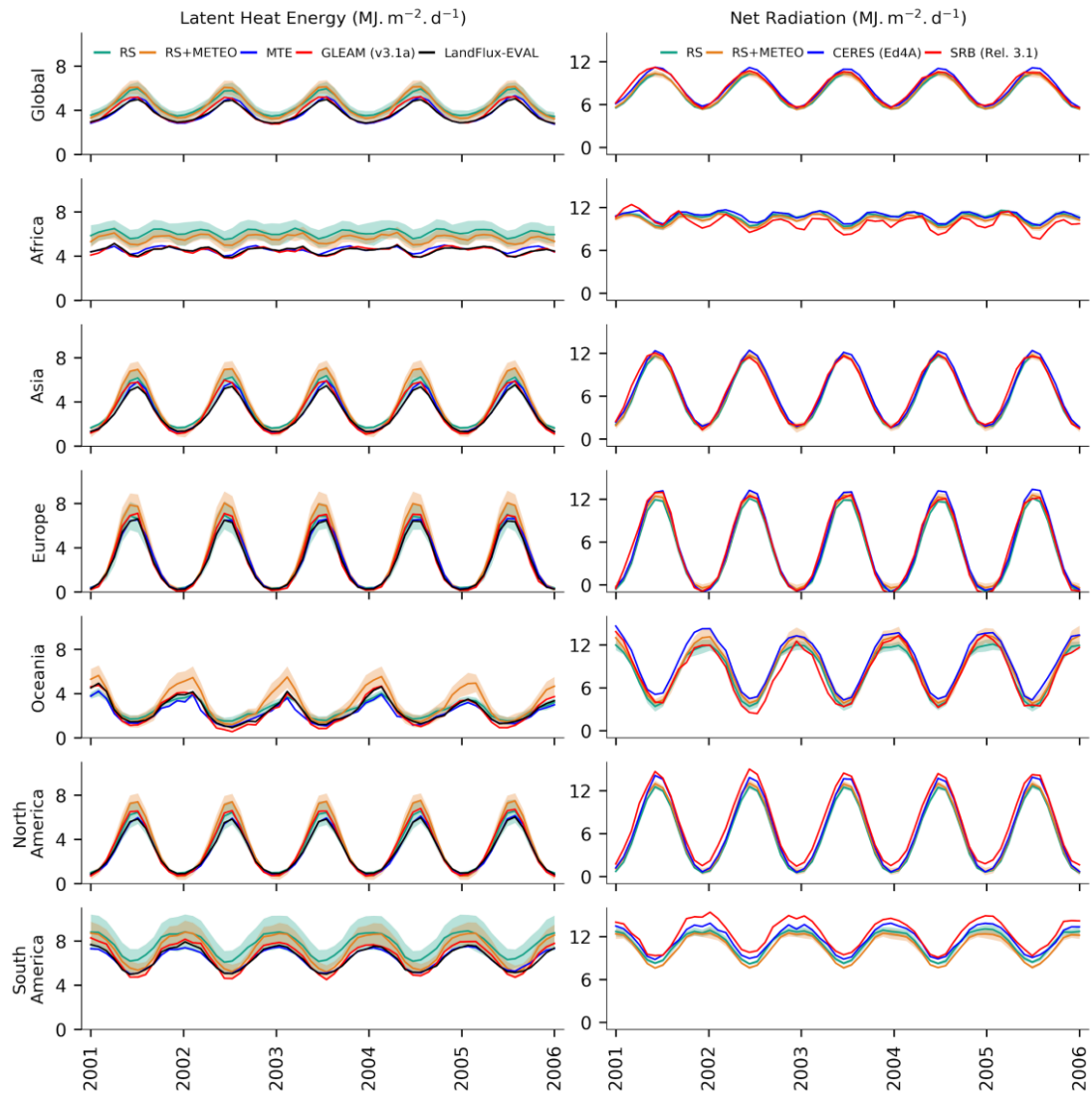

**Figure 6. Comparison of the temporal variations of global and continental latent heat energy (LE) and net radiation fluxes (Rn) from FLUXCOM with previous estimates.** The shaded region across the RS+METEO and RS lines indicate the uncertainty (± 1 robust standard deviation) of ensemble members. Note the different axes limits for LE and Rn.

We further compare the monthly variations of global and continental scale energy fluxes (Figure 6) against previous estimates. There is very high agreement among all with respect to seasonality. In all continents except Africa, the previous estimates are within the 1 standard deviation of RS and RS+METEO FLUXCOM products. In Africa, LE is higher in FLUXCOM products, which contributes to the slightly larger global ET in FLUXCOM than previous estimates. For Rn, the differences are relatively smaller in all continents and uncertainties obtained from the FLUXCOM ensemble are small compared to those of LE and H. The uncertainty estimates of LE in both RS and RS+METEO energy fluxes show distinct seasonal variation in all continents. In Africa and South America, the uncertainty ranges are larger in all seasons. In other continents, the uncertainty ranges are larger in the peak season and generally scale with flux magnitude.

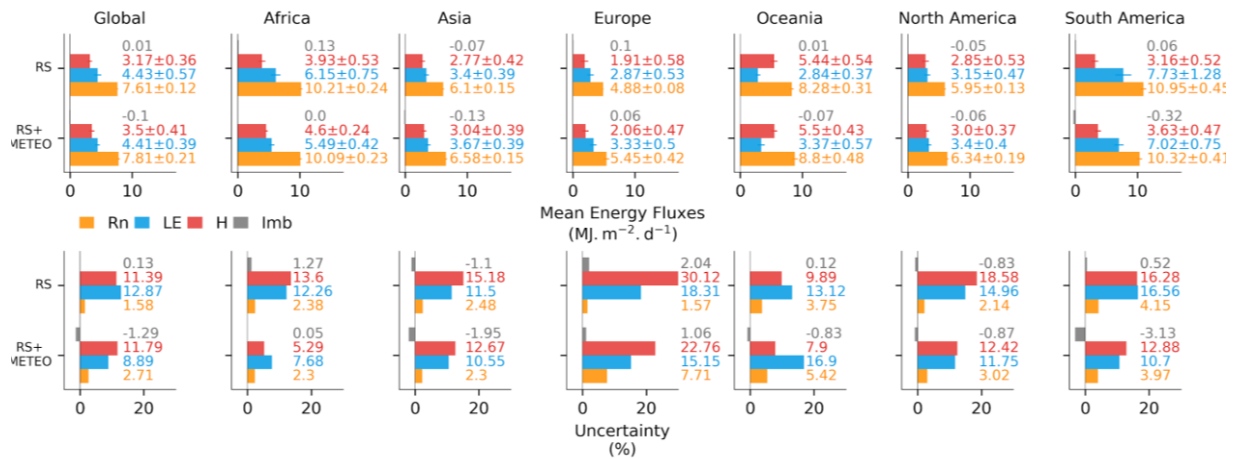

**Figure 7. Overview of the global and continental estimates of energy fluxes from FLUXCOM with uncertainties.** Top: Bars represent ensemble median of long-term means (2001-2013) of ensemble members for net radiation (Rn, orange), latent heat (LE, blue), sensible heat (H, red), and the energy imbalance (Imb, Rn-LE-H, grey) and error bars refer to one robust standard deviation (not available for Imb). Bottom: Bars represent uncertainties relative to the ensemble median in %; the relative uncertainty of Imb is (Rn-LE-H)/Rn. Note that all estimates refer to the global vegetated area.

We have further summarized mean annual energy budgets with uncertainties for the vegetated area of the globe and over all continents (Figure 7, top panel). The global and continental energy budgets of the RS and RS+METEO products are consistent with each other. Globally, both RS and RS+METEO products show that most of Rn is partitioned to LE. Only Oceania (dominated by Australia) shows more sensible than latent heat. In Africa LE is significantly larger than H due to the exclusion of the non-vegetated Sahara Desert, where H is expected to be much larger than LE.

The mean annual imbalance, defined as Rn-LE-H is close to zero. This indicates that the tower-to-globe scaling across all energy balance correction variants is robust and did not introduce sizeable biases. Inspecting relative uncertainties (Figure 7 lower panel) we find that Rn is typically constrained by less than 5%, uncertainties for latent and sensible heat fluxes are on the order of 10-20%.

For comparisons with published global estimates we scaled FLUXCOM values using estimates of H and Rn for hot and cold deserts (see Methods). We obtained mean global values of H of 2.85 MJ m$^{-2}$ day$^{-1}$ for the RS products and 3.10 MJ m$^{-2}$ day$^{-1}$ for the RS+METEO products, or, equivalently, 33 ± 4 and 35.9 ± 5 W m$^{-2}$ (uncertainties taken from Figure 7) respectively. These values are larger than the 27 W m$^{-2}$ reported by Trenberth et al.[43] and somewhat smaller than the range of 36-40 W m$^{-2}$ given by Siemann et al.[42]. FLUXCOM estimates are however in good agreement with the values of Wild et al.[4] of 32 W m$^{-2}$ as a best estimate derived as energy budget residual of observational data and the value of 38 ± 6 W m$^{-2}$ estimated by L'Ecuyer et al.[3] based on constraining the global energy and water cycles by multiple data streams. Scaled FLUXCOM Rn (see Methods) yields 75.8 ± 1.4 W m$^{-2}$ and 77.6 ± 2.4 W m$^{-2}$ for RS and RS+METEO respectively which is in excellent agreement with the best estimate of L'Ecuyer et al.[3] of 76 W m$^{-2}$.

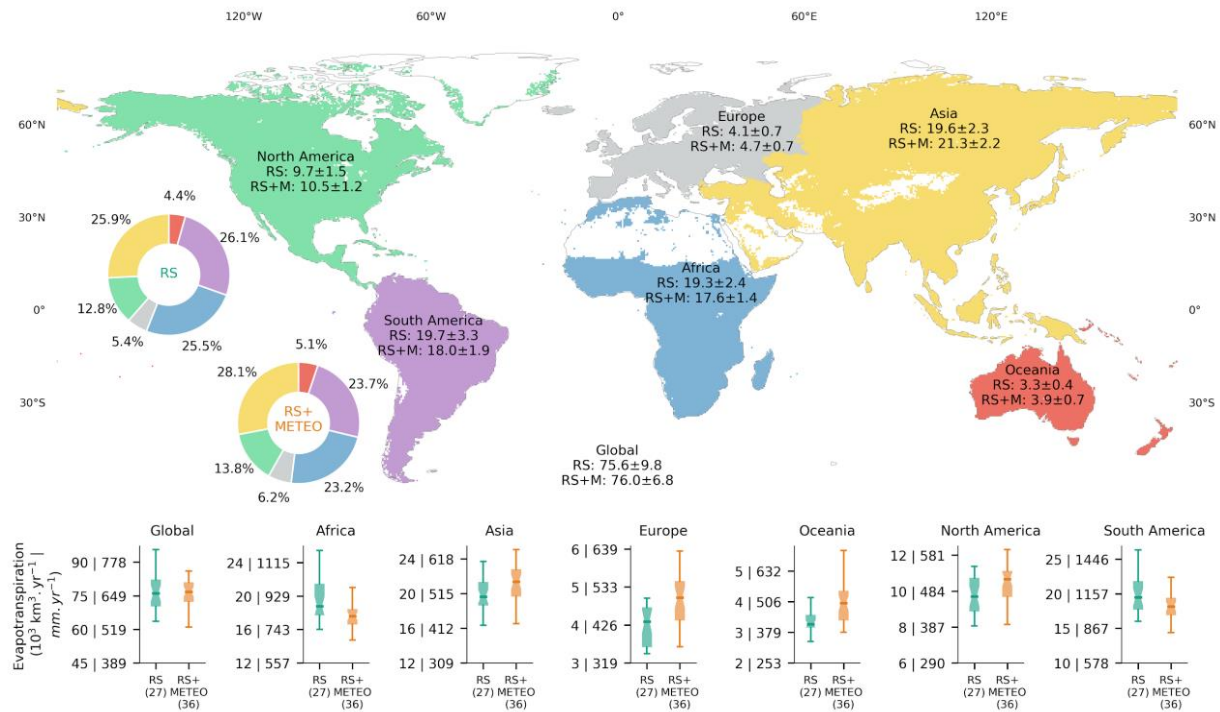

**Figure 8. Overview of the global and continental evapotranspiration (*ET*) fluxes from FLUXCOM.** Numbers in the map refer to the median +/- 1 robust standard deviation of RS and RS+METEO (RS+M) ensemble members in *1,000 km³ yr⁻¹*. Donut charts show the the relative contributions of each continent to the global *ET* with colours corresponding to the map. Box plots show the ensemble spread of the 27 and 36 members for RS and RS+METEO respectively, where the whiskers indicate the 1.5 times interquartile range. Note that the y-axis scale changes between continents, and that labels for ET in *mm yr⁻¹* are with respect to the vegetated area.

Because latent heat, as evapotranspiration (ET), is also a critical component of the water cycle, we summarize the continental and global ET from FLUXCOM (Figure 8). Globally, the ET from RS and RS+METEO are 75.6±9.8 and 76±6.3 ×10³ km³ yr⁻¹, respectively. These global ET are at the upper end of previously reported values (65-75). Several studies[9,44-46] indicated global ET in the range of 65 to 70 ×10³ km³ yr⁻¹. More recently, global ET values in the range of 70-75 ×10³ km³ yr⁻¹ were reported[47-49]. Interestingly, global ET estimated from an energy balance perspective also tend to yield values at the upper end of commonly reported values such as Trenberth[43] (73 ×10³ km³ yr⁻¹), L'Ecuyer[3] (72±5 ×10³ km³ yr⁻¹), and Wild[4] (72 within a range of 64 – 85 ×10³ km³ yr⁻¹).

Around three quarters of the global ET is equally distributed across Africa (23.2 – 25.5%), Asia (25.9 – 28.1%) and South America (23.7-26.1%), while the lowest contribution is from Oceania (4.4-5.1%), in both RS and RS+METEO (see donut charts in Figure 8). Both RS and RS+METEO show sizeable uncertainty ranges of global and continental ET (see box and whisker plots in Figure 8). In all continents, the ensemble medians of RS and RS+METEO overlap. In Asia, Europe, Oceania, and North America, RS+METEO has larger uncertainties than RS, while the opposite is true in Africa and South America.

## Usage Notes

For cross-consistency analysis and evaluation of LSM simulations we suggest focusing on spatial patterns of mean annual and mean seasonal fluxes. For comparison with offline LSM simulations we recommend using products from the RS+METEO setup forced with corresponding meteorological forcing to minimize deviations due to different climate input data. However, as RS+METEO inputs prescribe seasonal and spatial land surface properties--through mean seasonal cycles and PFT-based tiling respectively--full consistency with forcing-specific LSM simulations cannot be achieved. For example, while fAPAR is prescribed by remote sensing covariates in FLUXCOM, it is simulated by the LSM from the climate forcing. Since products from the RS setup are not subject to uncertain meteorological inputs the RS products may be preferable for energy and water budget studies or for evaluating the choice of climate input driver on LSM simulations.

Patterns of interannual variations in these products are expected to be more uncertain than spatial patterns of mean annual or seasonal fluxes. Experiences from FLUXCOM carbon fluxes suggest that magnitudes of interannual variations[50] are also likely too small in the energy flux products, and a normalization of the monthly or annual anomalies is recommended when comparing for example with LSM simulations. For example, global grids of LSM and FLUXCOM anomalies could be normalized by their standard deviations of the globally integrated anomaly time series to preserve spatial temporal patterns but to remove differences in variance. Note that interannual variations in the RS+METEO products originate exclusively from direct effects of changing weather with remotely sensed surface properties being constant between years. This is particularly important for studies on phenology or the impact of land surface changes on land-atmosphere energy fluxes where we would recommend using RS setup products. Low frequency variations and trends require very cautious interpretation as factors expected to cause trends, most importantly the physiological effects of rising $CO_2$, are not accounted for. Trends in land surface properties such as greening or browning are not accounted for in the RS+METEO setup since mean seasonal cycles of MODIS land products were used here. In comparison, the RS products do have these trends included but due to issues with sensor age-based drift in MODIS reflectances caution is warranted.

The energy flux densities of FLUXCOM product are defined per vegetated area in each grid cell. This needs to be considered when calculating global and continental budgets, in particular for sensible heat and net radiation where these fluxes have sizeable magnitudes over non-vegetated areas (see Technical validation). The land fraction provided with the FLUXCOM data should be multiplied with the flux densities for a correct accounting of fluxes over land.

FLUXCOM ensemble products provide the median absolute deviation of ensemble members per grid cell and time step which might be scaled to a robust estimate of the standard deviation of a normal distribution by multiplying by 1.4826. A propagation of this spatially and temporally explicit uncertainty to a temporal aggregated (e.g. mean annual) or spatial (e.g. continental) uncertainty would require assumptions on error co-variances in space and time. In such cases, we recommend to perform the desired aggregation for each ensemble member separately and subsequently take the spread of the aggregated ensemble members as the uncertainty metric. If users require a different combination of ensemble members other than those presented here, have questions or want to give feedback, please contact Martin Jung (mjung@bgc-jena.mpg.de).

## Acknowledgements

GCV. was supported by the European Research Council (ERC) under the ERC Consolidator Grant ERC-CoG-2014 SEDAL (grant agreement 647423). GT was supported by the BACI project funded by the EU's Horizon 2020 Research and Innovation Programme (GA 640176).

## Author contributions
MJ coordinated the FLUXCOM initiative, designed the study, and drafted the manuscript. SK analysed the results and produced the figures. UW helped with data processing. FG helped with coding of data processing. KI contributed the processing of satellite data and with machine learning code. DP, GT,

CS, and GCV contributed machine learning code. MJ, SK, KI, DP, GT, CS, GCV, and MR contributed intellectual input, and edited the manuscript.

## Competing interests
The authors declare no competing financial interests.

Manuscript submitted to Scientific Data on 10th December 2018

## Data Citations

20